\documentclass[final,authoryear,3p,times,twocolumn]{elsarticle}

\usepackage{amssymb}

\newcommand\integral{\textsl{INTEGRAL}}
\newcommand\swift{\textsl{Swift}}

\journal{Astronomy and Computing}

\begin{document}

\begin{frontmatter}



\title{GOLIA: an \textit{INTEGRAL} archive at INAF-IASF Milano}


\author[label1]{A. Paizis}
\ead{ada@iasf-milano.inaf.it}
\author[label1]{S. Mereghetti}
\author[label2]{D. G\"{o}tz}
\author[label1]{M. Fiorini}
\author[label3]{M. Gaber}
\author[label1]{R. Regni Ponzeveroni}
\author[label1]{L. Sidoli}
\author[label4]{S. Vercellone}

\address[label1]{INAF-IASF, Sezione di Milano, Via Bassini 15, 20133 Milano, Italy}
\address[label2]{AIM/Service d'Astrophysique - CEA Saclay DSM/Irfu, Gif sur Yvette, France}
\address[label3]{Former ISDC, Chemin d'Ecogia 16, 1290 Versoix, Switzerland}
\address[label4]{INAF-IASF, Sezione di Palermo, Via Ugo La Malfa 153, 90146 Palermo, Italy}

\begin{abstract}
We present the archive of the {\it INTEGRAL} data developed and maintained at INAF-IASF Milano. 
The archive comprises all the public data currently available (revolutions 0026-1079, i.e., December 2002 -- August 2011). 
{\it INTEGRAL} data are downloaded from the ISDC Data Centre for Astrophysics, Geneva, on a regular basis as they become public and a customized 
analysis using the OSA 9.0 software package is routinely performed on the IBIS/ISGRI data. The scientific 
products include individual pointing images and the associated detected source lists in the 17--30, 30--50, 
17--50 and 50--100\,keV energy bands, as well as light-curves binned over 100\,s in the 17--30\,keV band for 
sources of interest. Dedicated scripts to handle such vast datasets and results have been 
developed. We make the analysis tools to build such an archive publicly available. The whole database (raw data and products) enables an easy access to the hard X-ray long-term 
behavior of a large sample of sources. 
\end{abstract}


\end{frontmatter}


\section{Introduction}
\label{Intro}

The INTErnational Gamma-Ray Astrophysics Laboratory, {\it INTEGRAL} \citep{winkler03}, is a medium sized ESA mission successfully launched on October 17th, 2002. Its payload consists of two main gamma-ray instruments, the spectrometer SPI \citep{vedrenne03} and the imager IBIS \citep{ubertini03}, covering the 15\,keV -- 10\,MeV band.  IBIS is a high angular  resolution gamma-ray imager optimized  for accurate point source imaging and for the continuum and broad line 
spectroscopy. It consists of two layers,  the lower energy one 
\cite[IBIS/ISGRI, 15\,keV -- 1\,MeV, key parameters in Table~\ref{tab:table1},][]{lebrun03} and the higher energy one \cite[IBIS/PICsIT, 0.175--10\,MeV,][]{labanti03}. 
Co-aligned with SPI and IBIS are two X--ray monitors JEM--X \cite[4--35\,keV,][]{lund03} and an optical monitor OMC \cite[500--600\,nm,][]{mashesse03}. \integral~nicely links the energy band of pointed soft X-ray missions (such as {\it Chandra, XMM-Newton}) to high energy gamma ray space missions such as AGILE, Fermi and ground based TeV observatories.

Operating since 2002 and thanks to the synergy of its complementary instruments, {\it INTEGRAL} provides an impressive  quantity of high quality data for the  study of key science areas, such as nucleosynthesis, diagnostics of supernovae and supernovae remnants, compact sources (black holes, neutron stars, white dwarfs), electron-positron annihilation emission, super-massive black holes, the cosmic X-ray background, gamma ray bursts, etc  \cite[see][and references therein for a recent review]{winkler11}. At the time of writing a total of about 8.5 years of data have become public and are available to the scientific community for exploitation.

\integral~instruments have very wide fields of view. Even considering the fully coded field of 
view, where the signal to noise ratio is maximal, that is smaller than the total telescope field of view,
   we obtain fields of about $9^{\circ}\times9^{\circ}$ for IBIS, $16^{\circ}$ (corner to corner) for SPI, 
   $5^{\circ}$(\O{}) for JEM--X and  $5^{\circ}\times5^{\circ}$ for OMC. Hence a complete study of a 
   given source (or class of sources) requires huge datasets to be downloaded each time. 

In order to increment and ease our exploitation of \integral~data, we undertook the task of preparing and maintaining an \integral~archive, GOLIA (Giant On-Line \integral~Archive). 
GOLIA's aims are twofold: provide a local database of the available public data to facilitate personalized analysis of already downloaded data, and offer easy-to-browse IBIS/ISGRI data products for a quick and efficient view of the hard X--ray sky. 
All of which is locally and interactively available at INAF-IASF Milano. The scripts we have used to build 
GOLIA are publicly available, as described in Section~\ref{sec:scripts}. 

\section{A walk through GOLIA}\label{sec:walk}

\begin{figure*}
\begin{center}
\includegraphics[width=1.0\linewidth]{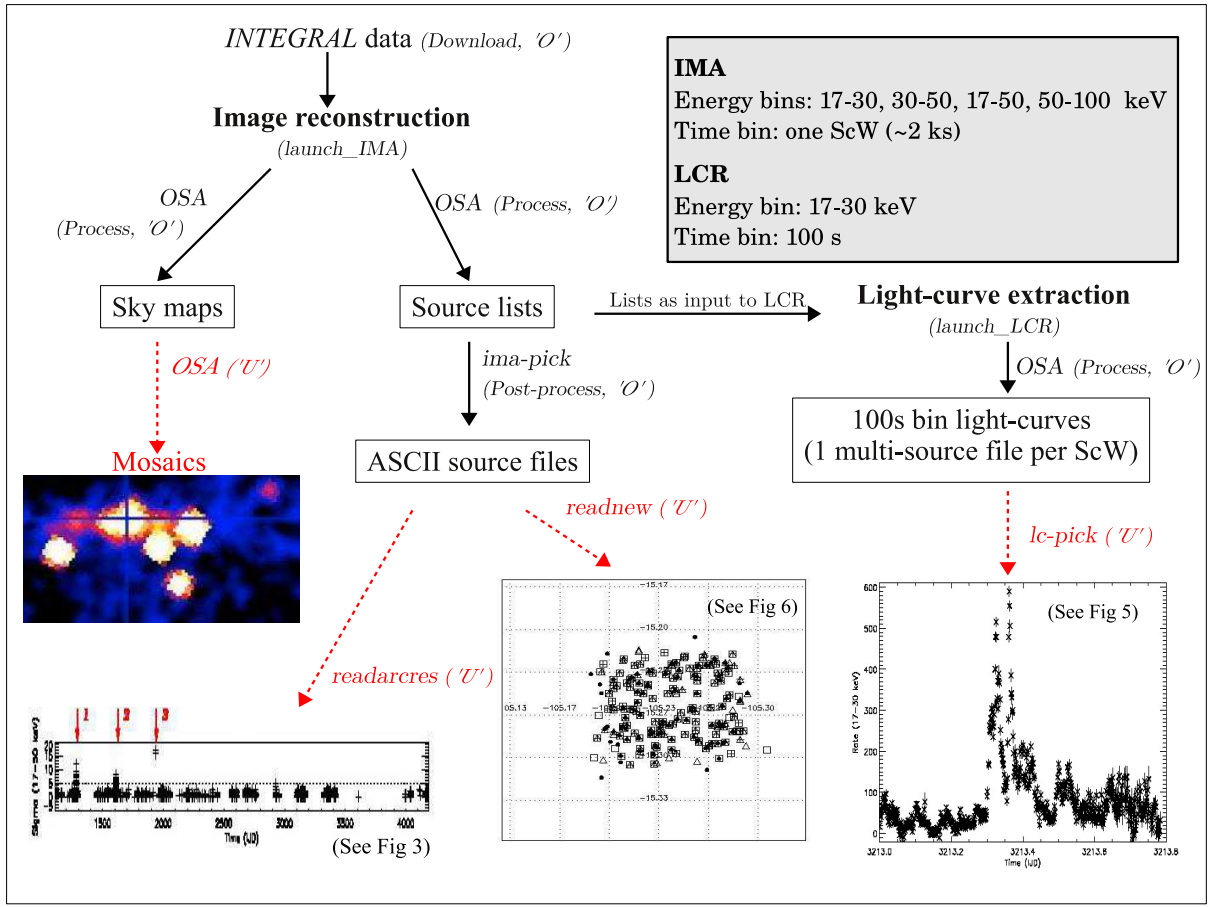}
\caption{Basic structure of GOLIA. The black solid arrows show the steps made by the archive \textit{Owner} (indicated as \textit{'O'}), 
while the ones shown with dashed red arrows are performed by the archive \textit{User} (indicated as \textit{'U'}). The Download, Process and Post-process steps are shown. All the scripts of these three phases are publicly available (see Section~\ref{sec:scripts}). The tools $\mathsf{launch\_IMA}$, $\mathsf{launch\_LCR}$,  $\mathsf{ima\_pick}$ , $\mathsf{readarcres}$ and $\mathsf{readnew}$ have been 
developed specifically for our archive. See Section~\ref{sec:walk}.}\label{fig:box}
\end{center}
\end{figure*}

The basic structure of GOLIA is shown in Fig.~\ref{fig:box} and is described below.

 \subsection{Building the archive (archive \textit{Owner})}\label{sec:golia}

\begin{table}
  \begin{center}
    \caption{Key parameters for IBIS/ISGRI  \citep{winkler03, winkler11}}
    \renewcommand{\arraystretch}{1.2}
    \begin{tabular}[h]{cc}
      \hline
      \hline
Parameter  & IBIS/ISGRI  \\
\hline
\hline
Energy range & $15$\,keV -- $1$\,MeV\\
Detectors  & $16384$ CdTe\\
Spectral resolution (FWHM) & $8$\,keV @ $100$\,keV\\
Field of view (fully coded) & $9^{\circ}\times9^{\circ}$\\
Angular resolution (FWHM) & $12^{\prime}$\\
Source location (radius) & $1^{\prime}$ for S/N=$30$\\
 &                        $3^{\prime}$ for S/N=$10$\\
Continuum sensit. @$100$\,keV &  $2.9\times$10$^{-6}$  ph cm$^{-2}$ s$^{-1}$ \\
 &  ($3\sigma$ detection in $10^5$\,s\\
  &  $\Delta$E=E/$2$)\\
Absolute timing accuracy ($3\sigma$) & $61$\,$\mu$s\\
\hline
\hline
      \end{tabular}
    \label{tab:table1}
  \end{center}
\end{table}

A detailed description of the \integral~Off-line Scientific Analysis (OSA) software package, 
used to analyze \integral~data, is
beyond the scope of the paper and we refer to the Documentation web page\footnote{http://www.isdc.unige.ch/integral/analysis\#Documentation} of the ISDC Data Centre for Astrophysics \citep[former \textit{INTEGRAL} Science Data Centre,][]{courvoisier03}. 
We recall here one key point: \integral~data are organized into the so-called \lq\lq Science
Windows\rq\rq~(ScWs, that for the purposes of this work can be considered as corresponding to a
pointing at a fixed sky position, $\sim$2\,ks). During the scientific analysis, all the
ScWs belonging to the same observation are grouped together to form an \lq\lq Observation
Group\rq\rq. The OSA package performs tasks in a modular way, i.e. it first applies  photon energy
correction (COR level) to all the ScWs belonging to the Observation Group, then it performs the
second step, the Good Time Interval computation (GTI) for all the ScWs, and so on. When the
analysis of a single observation is to be done (e.g. a few days / weeks of data) this approach is
efficient since it simplifies tasks such as mosaicking or light-curve collection. But since a
global analysis of all the public data means handling several tens of thousands of pointings
(currently 83733) we chose a single ScW based approach: each Observation Group consists of a single ScW and the analysis loops over all the available ScWs. Basically, the analysis is run at the level of the smallest available \lq\lq data entity\rq\rq~(the ScW), with no a-priori grouping chosen by the archive  \textit{Owner}. A-posteriori grouping of ScWs, for mosaicking and light-curve collection, will then be possible as chosen by each archive \textit{User}.
Furthermore, if the analysis of one single ScW happens to fail (for any given reason, missing data, no GTI available, etc), the analysis will continue and be completed for the remaining ScWs, independently. The implementation of such a script can be done in many ways, in our case Shell scripts with a simple \lq\lq $\mathsf{foreach}$\rq\rq~loop have been chosen.\\

Our IBIS/ISGRI  processing and post-processing steps (all performed by the archive \textit{Owner}) are described below. All the scripts described in this section are publicly available, see 
Section~\ref{sec:scripts}.
\begin{enumerate}

\item{\textbf{Download}:  all the \integral~public data are downloaded via the ISDC dedicated web page\footnote{http://www.isdc.unige.ch/integral/archive\#DataRelease}.  }\\

\item{\textbf{Process}: the standard OSA 9 software package\footnote{The latest version of the software package can be found at http://www.isdc.unige.ch/integral/analysis\#Software} is run on all public data of the IBIS/ISGRI instrument, looping over ScWs via the Shell scripts  $\mathsf{launch\_IMA.sh}$ and $\mathsf{launch\_LCR.sh}$. We decided to run the image analysis (IMA) on the whole archive first, to have a quantitative view of the results, and then run the light-curve extraction (LCR) part on it. The scripts can however be merged to perform IMA \textit{and} LCR for each ScW. The following steps are performed:\\

\emph{2-1 Image reconstruction (IMA)}: for each pointing a standard analysis is performed in four energy bands: 17--30, 30--50, 17--50 and 50-100\,keV.
The results of this analysis are the starting point for pointing-based, $\sim$2\,ks, light-curves (for all sources of the input catalogue) and 
 maps/mosaics for a deeper analysis.\\

\emph{2-2 Catalogue creation}: for each ScW, a catalogue is created merging the bright sources discovered from the previous IMA step (with detection significance higher than 3) and sources of interest, \textit{regardless} their detection level, be they new or already known. This allows to detect short bursts from interesting sources that would be otherwise missed in a single pointing ($\sim$2\,ks). Successful tests have been made on known GRBs.\\

\emph{2-3 Light-curve extraction (LCR)}: 17--30\,keV standard 100\,s bin light-curve extraction is performed for all the sources of each  ScW catalogue, i.e. thanks to step (2-2),  also including sources of interest not necessarily detected in the $\sim$2\,ks image (step 2-1). 
\\

\emph{2-4 Clean and close down}: relevant files are zipped and intermediate levels canceled.}\\

\item{\textbf{Post-process}: 
for each pointing, the IMA analysis step produces images in several bands and lists ($\mathsf{FITS}$ files) with  position,  flux and significance of the sources. Since map and source list files are derived for each ScW, all the information about a given source is spread over many ScWs. The OSA~9 tool $\mathsf{src\_collect}$ permits to collect IMA source results into a single table. However, it enables to retrieve results  for one source at a time only and the results are saved in a  $\mathsf{FITS}$ table.  We have developed a  $\mathsf{PERL}$ script ($\mathsf{ima\_pick.pl}$) that extracts in one go single $\mathsf{ASCII}$ files for {\it all} the sources present in the archive, one file per source.  The tool can be run in a cumulative way (the $\mathsf{ASCII}$ files can be updated with the new incoming results). Each $\mathsf{ASCII}$ file contains the results of the imaging step
 and is the starting point for quick visualization by the \textit{Users}.
 
We note that the 
$\mathsf{ima\_pick}$ script collects results based on the source name, creating a
$\mathsf{SourceName.dat}$ file. Every time the analysis software (OSA) detects a new source, the name {\sl NEW\_1} is given (with incremental number for more new sources), hence in the whole catalogue we obtain several {\sl NEW\_1, NEW\_2}, etc.,  sources that are \emph{not} the same, but have different positions. All these  sources will be collected in the same $\mathsf{ASCII}$ file by $\mathsf{ima\_pick}$ and need to be disentangled based on their coordinates.

The introduction of the {\bf Post-process} step, via the $\mathsf{ima\_pick}$ tool (not part of the OSA  package), leads to a huge simplification in the visualization of the image products. The produced $\mathsf{ASCII}$ files are a very easy-to-handle starting point and indeed  whole-archive (8.5 year) light-curves can be extracted in matter of seconds for whichever source originally in the catalogue.
}
\end{enumerate}

\begin{figure}
\includegraphics[width=0.9\linewidth]{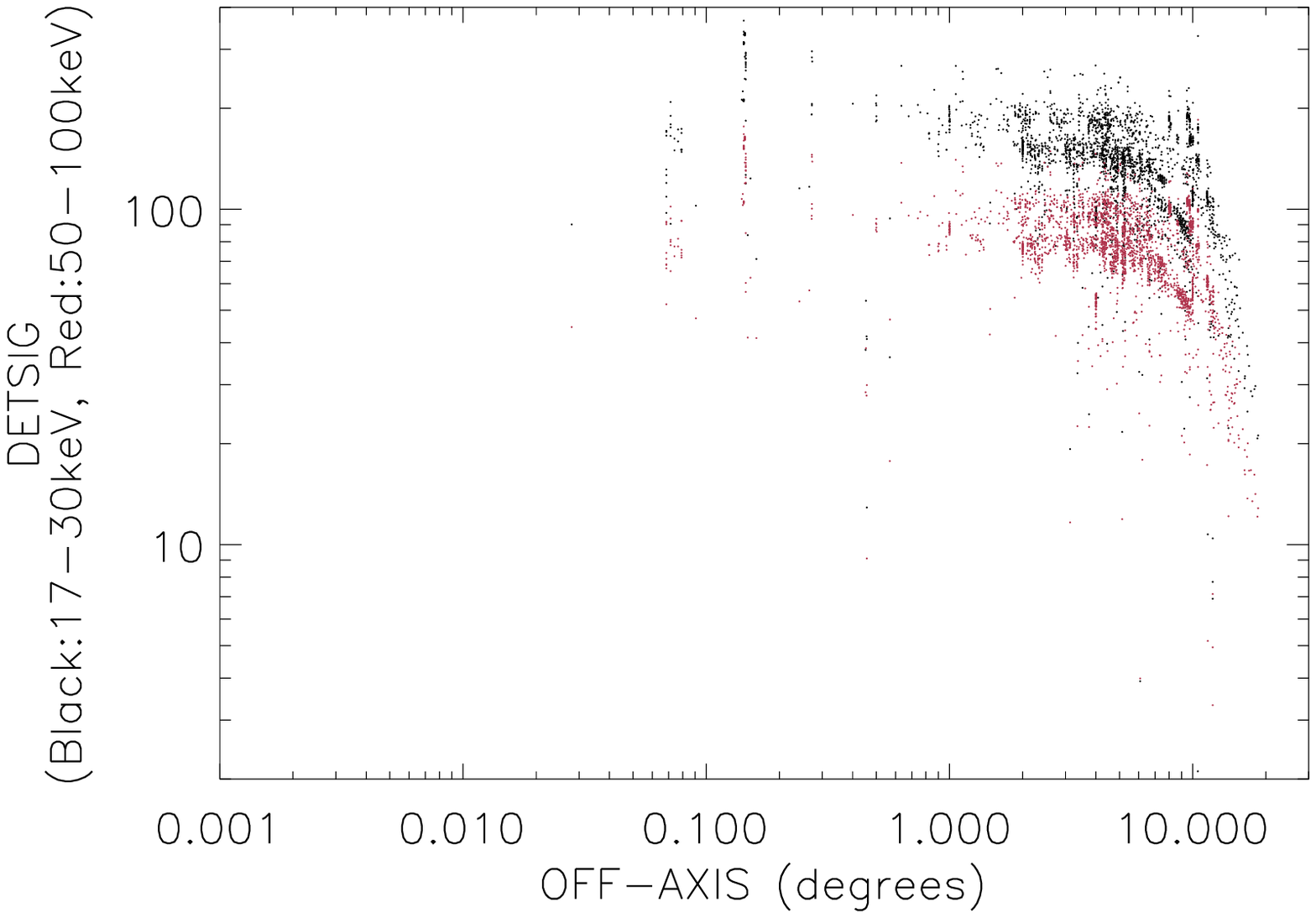}
\includegraphics[width=0.9\linewidth]{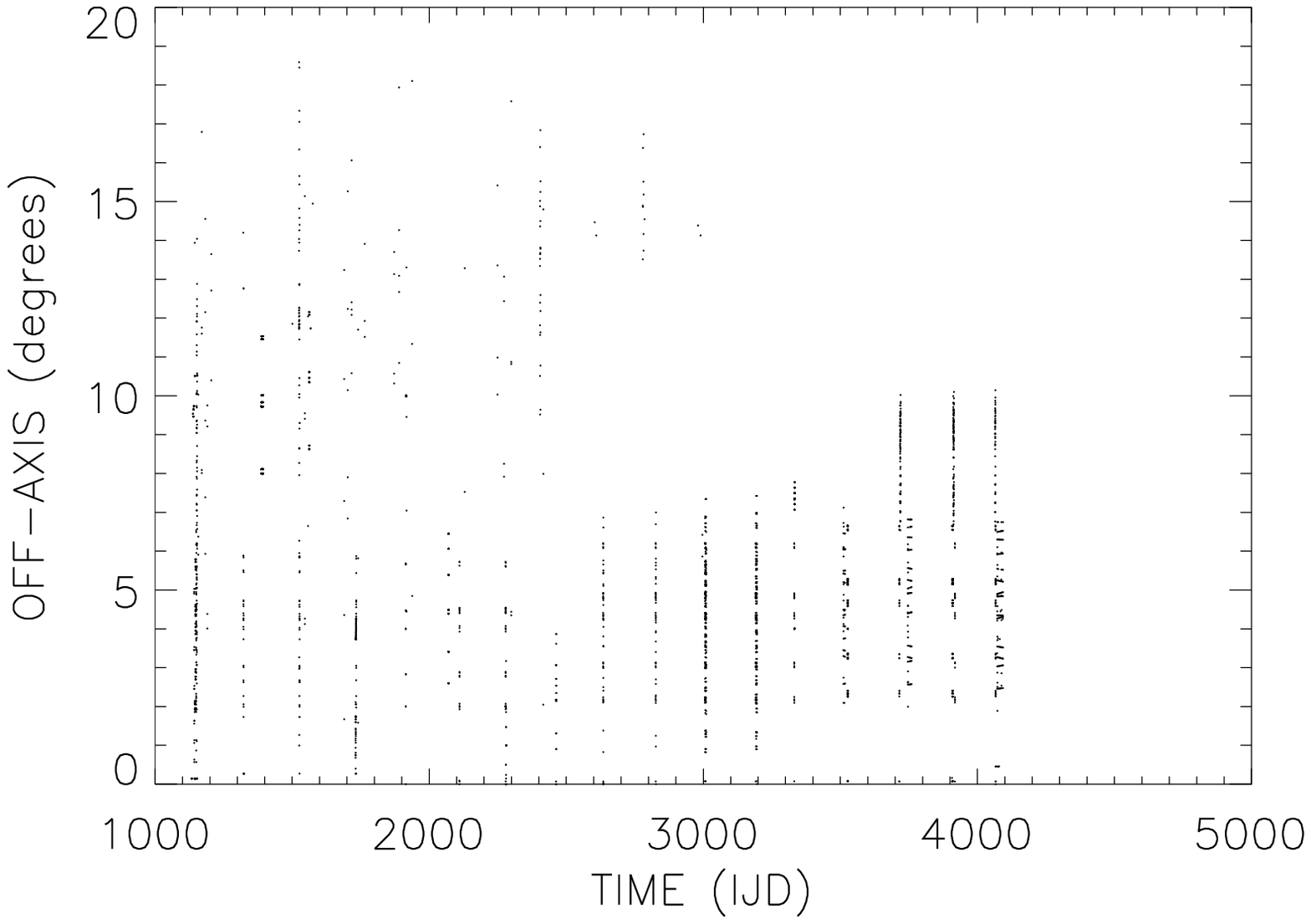}
\caption{Crab detection significance and off-axis angle. Each point is a single ScW, for a total of 2713 ScWs with the Crab in the IBIS field of view (off axis angle $<$20$^{\circ}$, exposure time of about 4\,Ms).}\label{fig:crab}
\end{figure}

 \subsection{Using the archive (archive \textit{User})}\label{sec:use}
Given the main steps performed by the archive \textit{Owner}~(Download, Process, Post-process), each \textit{User} can access GOLIA in many ways. 
The mere {\bf Download}, done by the \textit{Owner}, allows the \textit{User} to access  the entire \integral~public data for a personalized analysis, avoiding multiple time and space consuming downloads. 

Once the \textit{Owner} has completed the {\bf Process} step, the \textit{User} can already attain  analyzed ScW images to build customized mosaics and source searches via  a deeper analysis e.g. aimed to dim population studies. For brighter sources (or sources of interest) overall  light-curves binned at 100\,s can be collected with the  ISDC tool  $\mathsf{lc\_pick}$ (see Fig.~\ref{fig:lcr} and Section~\ref{sec:rapid}). The stacking of the \textit{Owner} ScW-based products of the \lq\lq Process\rq\rq~step (be they mosaics or 100\,s bin light-curve collection) can be time consuming for the \textit{User} if a very large number of ScWs is involved, as in standard OSA usage.

Finally, 
once all the key IMA parameters / results are saved in the $\mathsf{ASCII}$ file via the {\bf Post-process} step, it is immediate to plot them in different ways to have a quantitative view of the database. As an example, in Fig.~\ref{fig:crab} we show the detection significance (DETSIG) of the Crab in the whole archive in two energy bands, as a function of the off-axis angle, and the evolution of the latter with time, expressed in \integral~Julian Date, IJD\footnote{IJD = MJD - 51544.}. 

\begin{figure}
\includegraphics[width=1.0\linewidth]{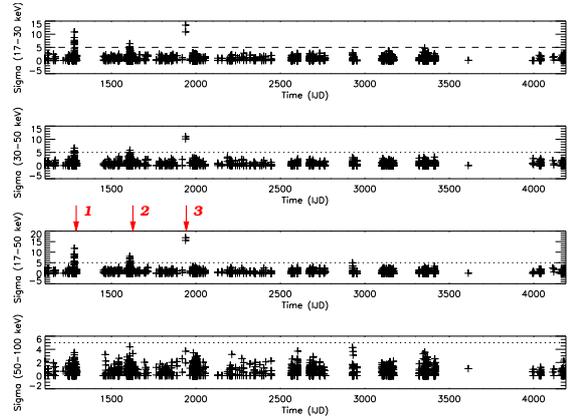}
\caption{Whole-archive $\sim$2\,ks light-curve obtained for IGR~J11215$-$5952. The plot shows all the \integral/IBIS public available data: 2442 pointings, 4.8\,Ms, from revolution 0030 (Jan 2003, IJD$\sim$1107) to 1060 (Jun 2011, IJD$\sim$4190).  Sigma $\geqq$5 (horizontal line) is a detection. }\label{fig:igr1121}
\end{figure}

ASCII files are easily readable / plottable by several software tools. In our case, we have developed two $\mathsf{IDL}$ scripts ($\mathsf{readarcres}$, $\mathsf{readnew}$) to read and plot such files. 
The former tool allows us to handle / plot single source $\mathsf{ASCII}$ files, providing long-term light-curves in terms of count-rate or detection significance (as in Fig.~\ref{fig:igr1121} for the detection significance case). 
The latter  ($\mathsf{readnew}$) allows us to investigate the new sources enabling the search of NEW sources by coordinates. This allows us to select the occurrences of each NEW source in the whole archive (see Fig.~\ref{fig:MAXI} and Section~\ref{sec:new}).

\section{GOLIA: a starting point for new investigations}\label{sec:results}
\begin{figure}
\includegraphics[width=0.6\linewidth,angle=270]{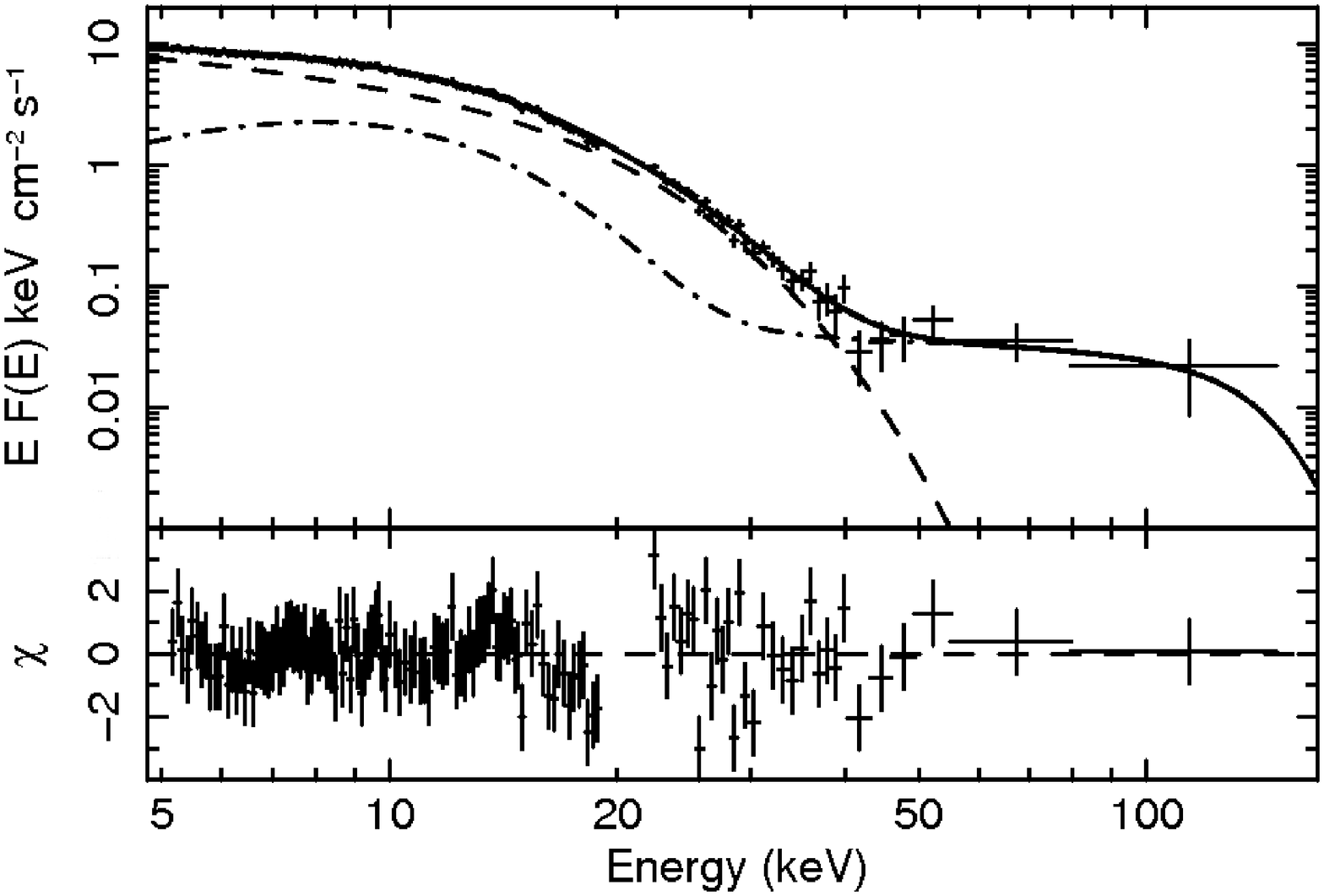}
\includegraphics[width=0.6\linewidth,angle=270]{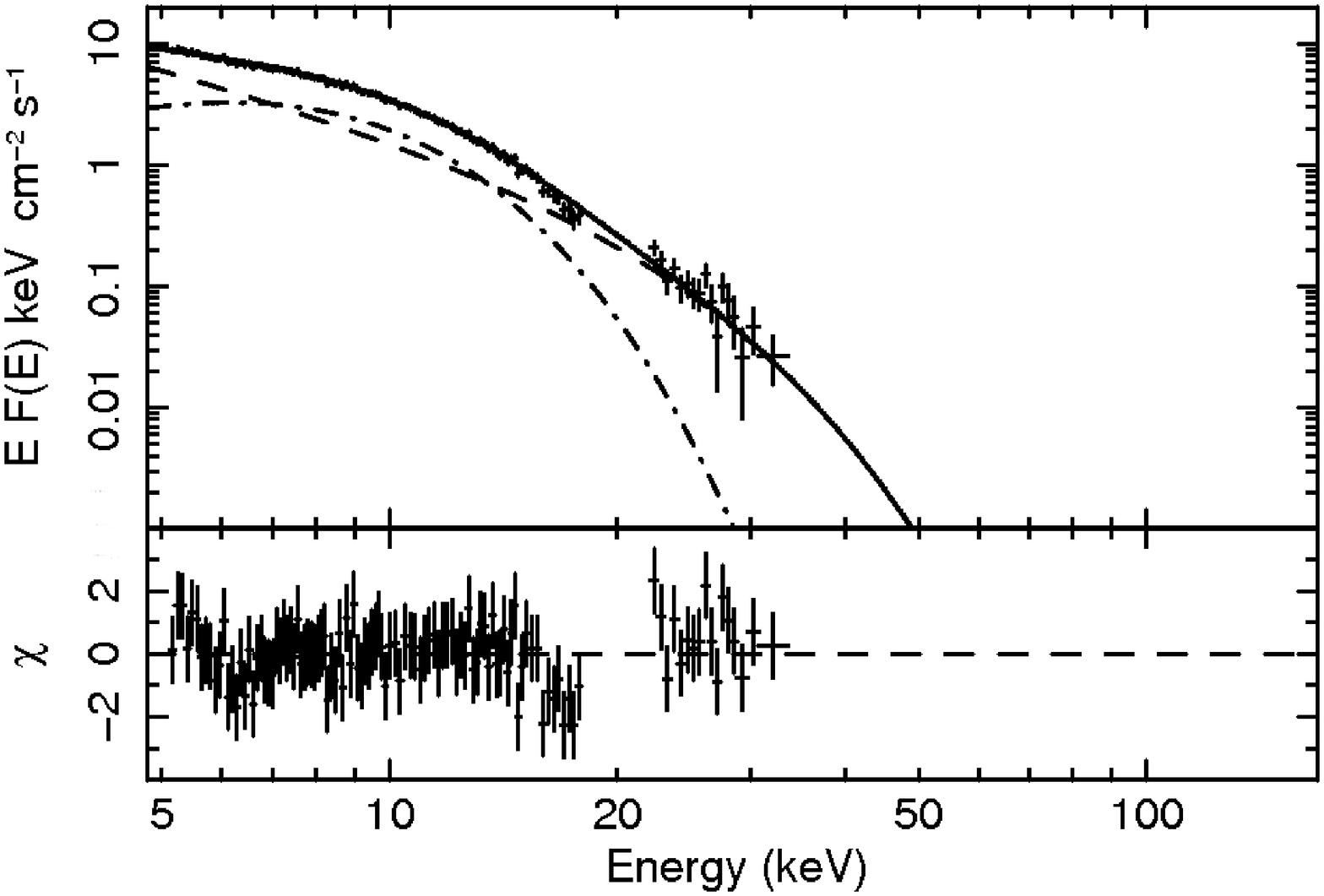}
\caption{ Unabsorbed EF(E) spectra with superposed best-fit  
$\mathsf{wabs\cdot(compTB[1]+compTB[2])}$  model 
 ({\it upper panels}) and residuals in units of $\sigma$ between data and
 model ({\it lower panels}) for GX~5--1.
 Different line styles represent
 single components of the best-fit model.  {\it Dashed-line}: pure thermal
 Comptonization for  $\mathsf{compTB[1]}$. {\it Dash-dot line}: 
  bulk+thermal Comptonization (upper spectrum) or simple BB (lower one)  for
$\mathsf{compTB[2]}$. {\it Solid line}: total spectrum \citep[from][]{mainardi10}. }\label{fig:tail}
\end{figure}
In this section we give examples of published and work-in-progress scientific results based on GOLIA.
\subsection{Supergiant Fast X-ray Transients (SFXTs)}\label{sec:sfxt}
Fig.~\ref{fig:igr1121} is an example of a whole-archive light-curve obtained with our interactive IDL $\mathsf{readarcres}$ script. The plot shows all the IBIS/ISGRI publicly available data of the transient \mbox{IGR~J11215$-$5952}.

With such a long-term view, outbursts, trends, periodicities can be easily spotted. Indeed, 
the automatic pipeline running on public {\it INTEGRAL} observations present in our archive
allowed us to discover that the Supergiant Fast X-ray Transient IGR~J11215$-$5952 displays 
periodic outbursts 
\citep{sidoli06}.
The source was discovered in 2005 \citep[arrow 3 in Fig.~\ref{fig:igr1121};][]{lubinski05}, but thanks to our systematic re-analysis of
{\it INTEGRAL} archival observations, we could discover two previously unnoticed outbursts (1 and 2 in Fig.~\ref{fig:igr1121}), spaced by intervals of 329
days, that occurred in July 2003 and May 2004, suggesting that this periodicity is the orbital period of the system.
At that time we could not exclude that the true periodicity is half of this period \citep[as indeed was later discovered, see][]{sidoli07}, because of the lack of {\it INTEGRAL} coverage of the source field of the first SFXT showing periodic outbursts.  

This discovery triggered by our archive has led to further investigations within our team, including the development of a stellar wind model for OB supergiants to explain the transient and variable nature of SFXTs, and its application to all public \integral~observations from 2003 to 2009 for a sample of 14 SFXTs \citep{ducci09,ducci10}.

\subsection{Low Mass X-ray Binaries (LMXB)}\label{sec:lmxb}
Visual inspection of light-curves such as Fig.~\ref{fig:igr1121} for persistently bright neutron star (NS) LMXBs (e.g. the Z source GX~5--1) has been the starting point for the systematic study of the evolution of the transient hard-tails in the spectra of these sources.  Indeed hard tails dominate the X-ray emission above 30\,keV, in an otherwise soft spectrum,  and they result in a source detection in the 30--50\,keV panel of Fig.~\ref{fig:igr1121}~\citep[e.g.][]{paizis06,farinelli08,mainardi10}. An example for the transient tail can be seen in Fig.~\ref{fig:tail} for GX~5--1 (JEM--X and IBIS/ISGRI data). The origin of such tails dominating the spectra above $\sim$30\,keV is still debated. \cite{paizis06} proposed for the first time a qualitative unified physical scenario to explain the spectral evolution of NS LMXBs, including the peculiar transient hard tail. 
As a following step, in the quest to study in a \emph{quantitative} way the evolution of the parameters describing the innermost physical conditions of NS LMXBs,  a new Comptonization model was developed \citep[$\mathsf{compTB}$\footnote{http://heasarc.gsfc.nasa.gov/docs/xanadu/xspec/models/comptb.html},][]{farinelli08}, 
 and successfully applied to a sample of bright NS LMXBs studied with \integral~\citep{mainardi10}.

\begin{figure}
\includegraphics[width=0.9\linewidth]{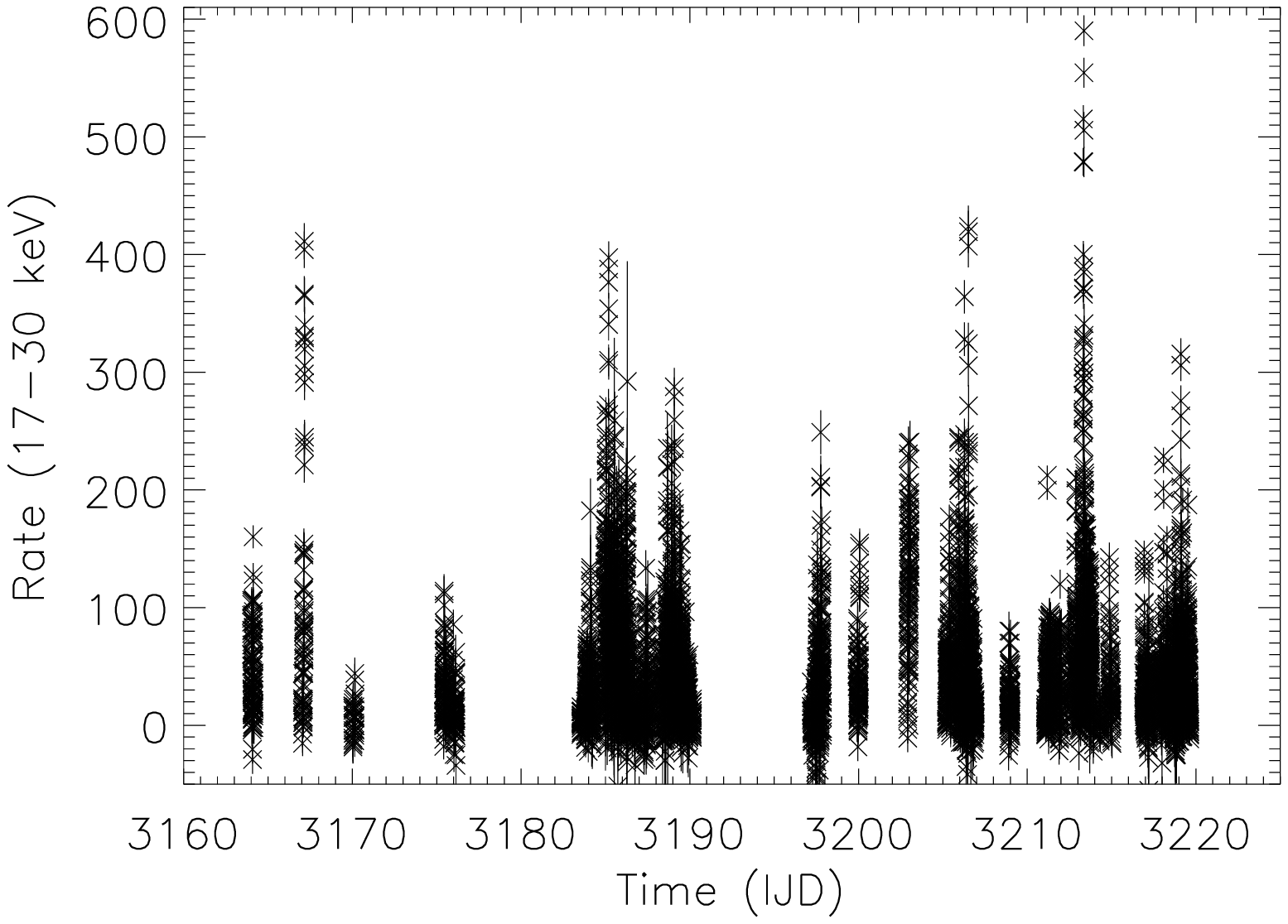}
\includegraphics[width=0.9\linewidth]{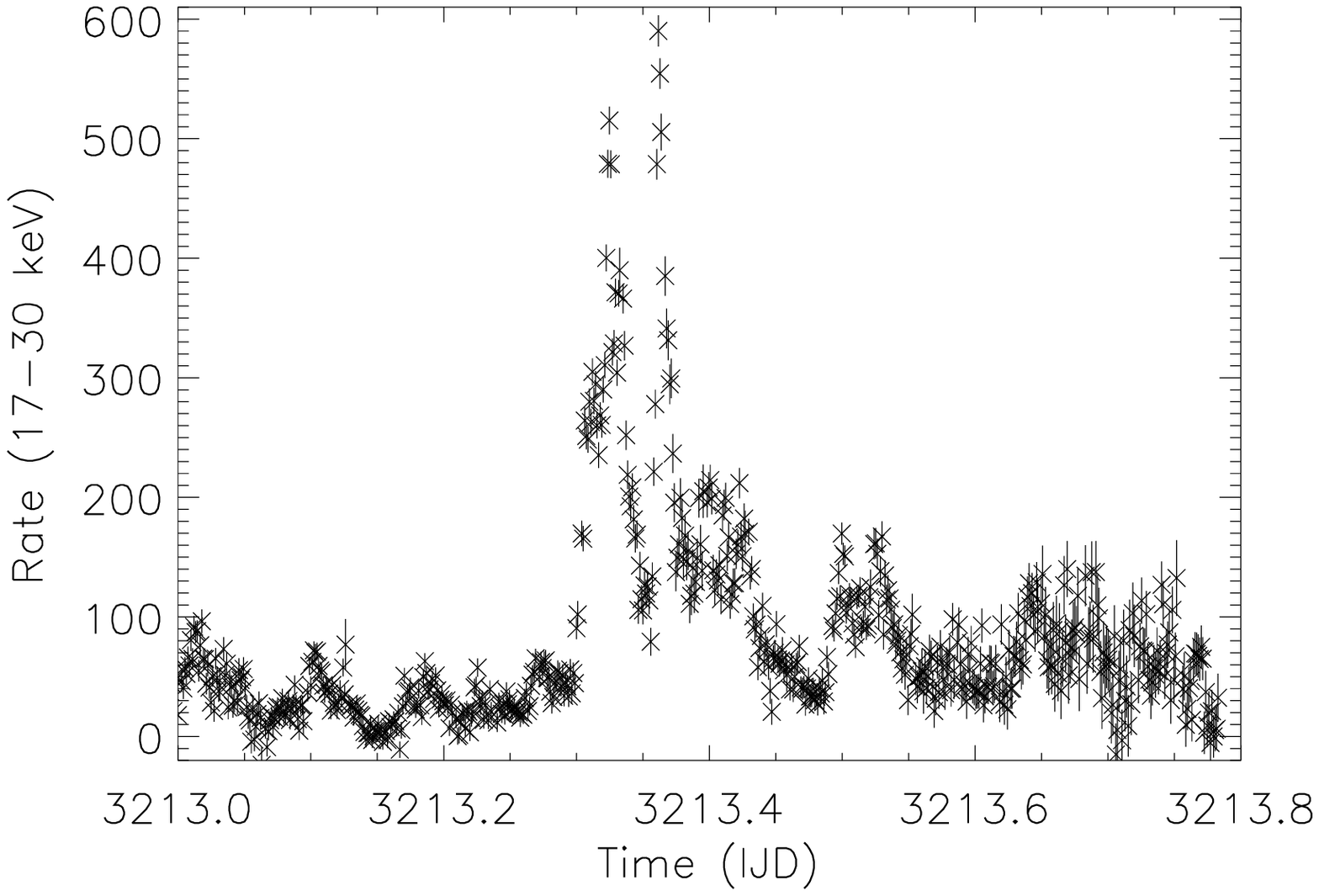}
\caption{17--30\,keV, 100\,s bin light-curve obtained for 4U~1700--377. \emph{Upper panel}: 2 month zoom.
\emph{Lower panel}:  0.8\,day zoom. 1\,Crab corresponds to a rate of about 173 counts/s in the same band.}\label{fig:lcr}
\end{figure}

\subsection{Rapid ($\sim$100\,s) variability}\label{sec:rapid}
In the archive, 100\,s bin light-curves in the 17--30\,keV band are available for all the sources detected in the
IMA part (2\,ks pointing basis) as well for sources of interest. Among others, we have extracted a whole-archive
light-curve for the high mass X-ray binary 4U~1700--377, for a total of 18\,Ms, sampled on a 100\,s basis. Two
different time zooms of the obtained light-curve (2\,months, 0.8\,days) are shown in Fig.~\ref{fig:lcr}. This complete light-curve is the starting point for a detailed study of the fast variability over 8.5 years of \integral~data in this peculiar source (Sidoli \& Paizis, in preparation).

\subsection{New Sources}\label{sec:new}
\begin{figure}
\includegraphics[width=0.9\linewidth]{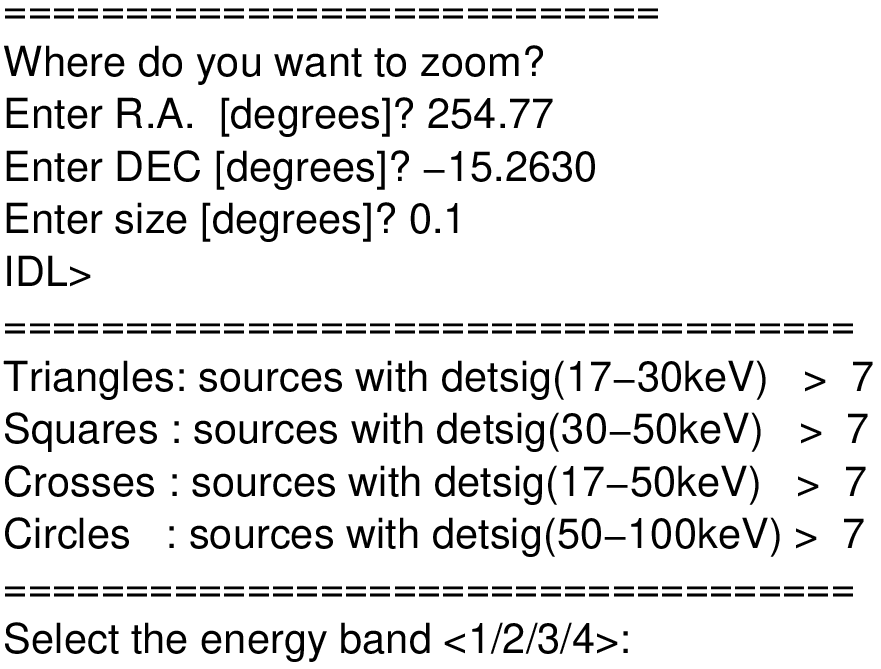}
\includegraphics[width=0.9\linewidth]{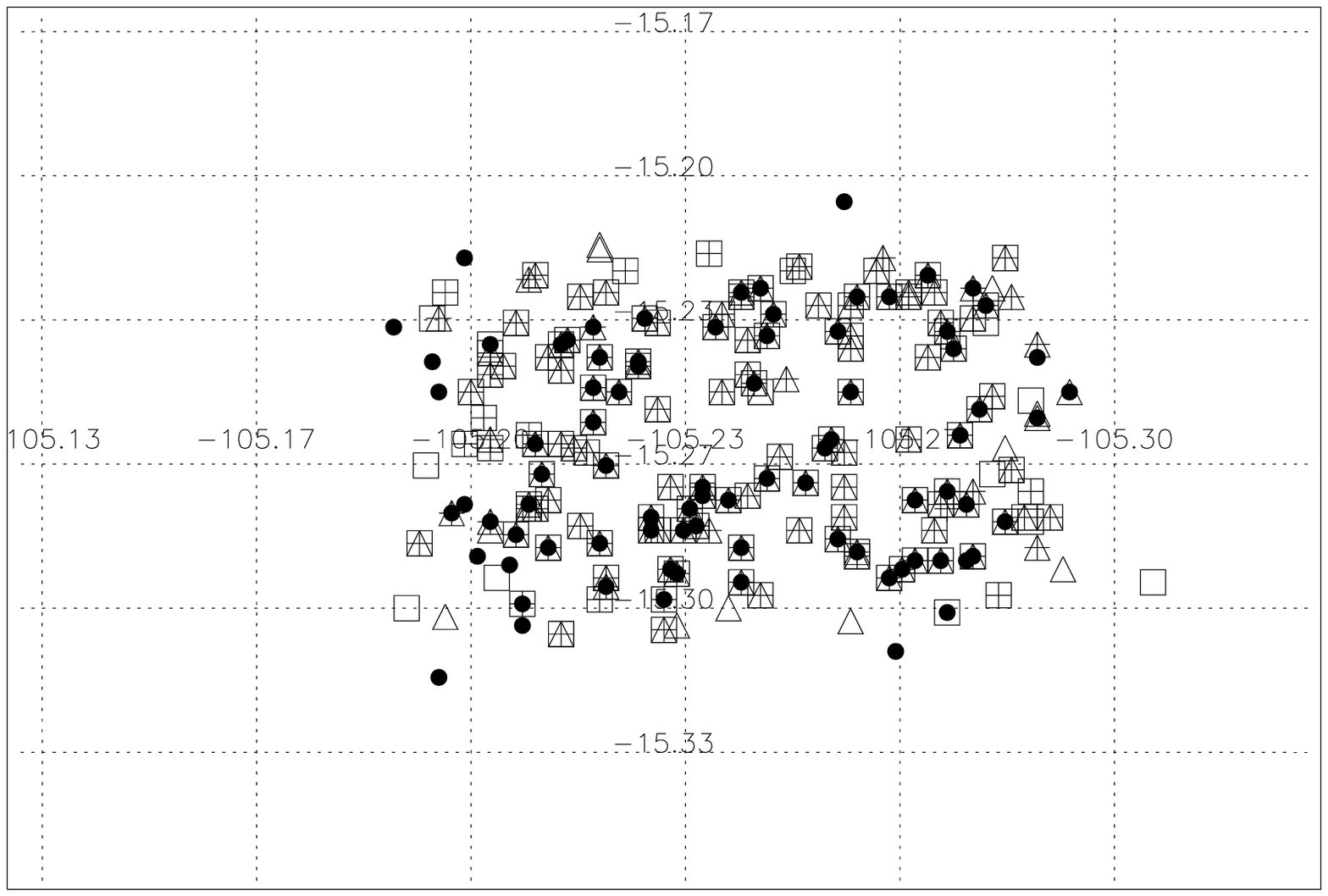}
\caption{Output of $\mathsf{readnew}$: on-screen text (upper panel) and visualization of multiple detections of the recently discovered transient MAXI~J1659$-$152 (lower panel). See Section~\ref{sec:new}.}\label{fig:MAXI}
\end{figure}
As discussed in Section~\ref{sec:use}, we have developed an interactive tool that reads the NEW source $\mathsf{ASCII}$ files created by $\mathsf{ima\_pick}$, grouping them by coordinates ($\mathsf{readnew}$). 

\begin{figure*}
\includegraphics[width=1.0\linewidth]{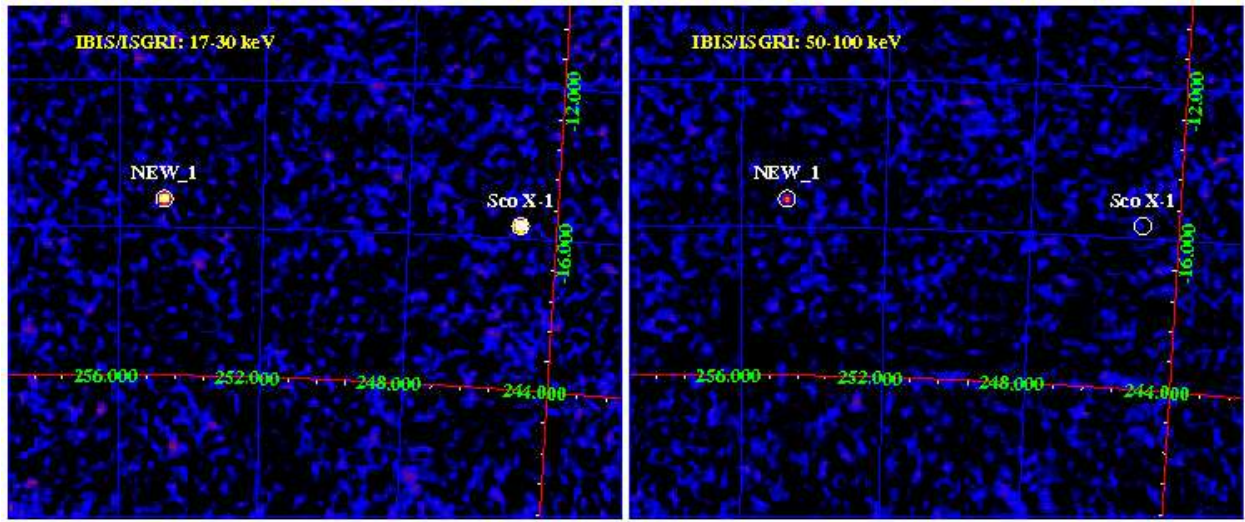}
\caption{IBIS/ISGRI sky maps of a single pointing (ScW \#092700700010) where MAXI~J1659$-$152 is detected by OSA and labeled as NEW (See Section~\ref{sec:new}).  }\label{fig:MAXI2}
\end{figure*}

Fig.~\ref{fig:MAXI}, lower panel, shows multiple detections (from different bands and pointings) of a new source, the recently discovered transient MAXI~J1659$-$152 \citep[a Black Hole candidate,][]{negoro10, negoro12}, not yet included in the \integral~General Reference Catalogue, hence labeled as new, NEW\_1, by the software. Fig.~\ref{fig:MAXI}, upper panel (text), shows some stages of the interactive process, with the description of the different symbols and region of the sky selected. A zoomed-in IBIS/ISGRI image of one of the multiple pointings of Fig.~\ref{fig:MAXI} can be seen in Fig.~\ref{fig:MAXI2} where  MAXI~J1659$-$152 is labelled as NEW\_1 by the software. The well known NS LMXB Sco~X--1, $\sim$9$^{\circ}$ apart, is also visible and appears much softer than MAXI~J1659$-$152: indeed, while in the softer band Sco~X--1 is the brightest source of the field (left panel, 17--30\,keV), in the harder band Sco~X--1 is not even detected (detection significance about 2), while MAXI~J1659$-$152 is still clearly there (detection significance about 10, 50--100\,keV, right panel). 

We are able to trace back in the \integral/IBIS archive  sources that were discovered recently (i.e. later than the creation of the \integral~General Reference Catalogue used, 2010 April 8th\footnote{We note that on 2012 November 15th, a new  catalogue has been delivered by the ISDC, including new sources reported in the literature: http://www.isdc.unige.ch/integral/science/catalogue.}), as well as to discover previously unnoticed NEW \integral~sources, such as IGR~J15283--4443, discovered by our systematic analysis of all \integral/IBIS public data \citep{paizis06b} and later on confirmed by \swift~follow-up observations \citep{rodriguez10}.

\section{Build your own GOLIA}\label{sec:build}

We report  the main characteristics of GOLIA (disk space, costs, time employed, etc.), in order to give a quantitative idea of the effort involved in such a work. The whole analysis has been performed on a single server HP ML330G6, openSUSE 11.4, two processors Xenon Quad-Core E5506, 2.13\,GHz and RAM 8\,GB (June 2011 cost $\sim$2300\,Euro). A total of 12 external disks of 1\,Tb each have been purchased to store the data and the results (213\,Euro each Tb). 
Currently a total of 83733 pointings have been analyzed, i.e. starting revolution 0026 (December 2002) to 1079 (August 2011), for a total of 8.5 years of data.  Disk space is needed for the raw data while the results themselves occupy a smaller fraction ($\sim$1.3\,Tb for 8.5 years of results versus about 4.8\,Tb of 8.5 years of data  -  excluding SPI raw data). 

The integrated IBIS/ISGRI analysis time of the 8.5 years of data has been about 8 months with one server (roughly half of which  for the IMA part and half for the LCR one). More servers (that can be used in parallel with GRID, the analysis script modular structure allows it) will result in a much lower processing time. 
The addition of further data products (more energy or time bins, spectral extraction or inclusion of other instruments) will of course increase the CPU time as well as disk space needed.

We have decided to begin extracting the results for IBIS/ISGRI  because we believe that it is a very good starting point to investigate the less explored hard X-ray sky (with respect to the softer JEM--X range), with a good (12$^{\prime}$ FWHM) angular resolution to disentangle sources in the Galactic plane and bulge (with respect to SPI,  2.5$^{\circ}$ FWHM). Furthermore, we have also been triggered by our experience in the IBIS/ISGRI data analysis that allowed us to handle large datasets while choosing few energy and time bins, still capable to unveil interesting source behaviour (softening versus hardening, unexpected variabilities, transient behaviour, etc) but yet within acceptable running time. 

Automatic spectral extraction, though applicable with our scripts, would have been extremely time consuming given our single-server setup. Furthermore, we decided not to perform it since the feedback with our potential users, and our previous experience on whole-archive spectra, suggested that source spectra should be re-extracted with optimized energy binning to be meaningful for a detailed analysis.

Once the setup is up and running, the maintenance of the archive is not an issue and neither is its regular update. Maps, light-curves, raw data, fits and $\mathsf{ASCII}$ files can all be accessed by the  users, enabling a personalized usage of the archive and triggering further detailed investigations such as the ones previously described.

\subsection{Our scripts}\label{sec:scripts}
All the \integral~public data can be freely downloaded via the ISDC dedicated web page (see footnote number 2).
A tar file with the scripts we used to build GOLIA (Process and Post-process steps, i.e. the ones performed by the archive \textit{Owner}, see Section~\ref{sec:golia} and Fig.~\ref{fig:box}) and a short description (README) can be downloaded from our GOLIA web page \footnote{http://www.iasf-milano.inaf.it/$\sim$ada/GOLIA.html}. These scripts are the Shell scripts $\mathsf{launch\_IMA.sh}$ (and its sub-routine $\mathsf{analysis\_IMA.csh}$),  
$\mathsf{launch\_LCR.sh}$ (and its sub-routine $\mathsf{analysis\_LCR.csh}$) used to launch OSA looping over ScWs, and the $\mathsf{PERL}$ script
$\mathsf{ima\_pick.pl}$, used to \lq \lq re-format \rq \rq~the IMA results. Together with OSA, these are the tools we used to build GOLIA, they use freely available interpreters and need no compilation. 

Once the archive is available, together with all the ASCII files created by $\mathsf{ima\_pick}$, browsing and visualization of the IMA results can be easily performed with any software that reads ASCII files. Mosaics of the images or collection of the 100s lightcurves can be done as described in the ISDC 
documentation.

\section{Conclusions}

 There are several on-going international collaborations and efforts to exploit and share \textit{INTEGRAL} data and results with the scientific community (to some of which we have participated as well). We recall here 
the Galactic bulge monitoring programme \citep{kuulkers07}\footnote{http://integral.esac.esa.int/BULGE/}, the \textit{INTEGRAL} Spiral Arms collaboration \footnote{ http://sprg.ssl.berkeley.edu/$\sim$bodaghee/isa/}, and the \textit{INTEGRAL} Galactic plane scan \footnote{http://gpsiasf.iasf-roma.inaf.it/}, all of which deal with  the analysis of IBIS/ISGRI and JEM--X data of specific sets of data (i.e. portions of the sky). Among the \textit{all} sky surveys we recall the IBIS/ISGRI catalogues by  \cite{krivonos07}\footnote{http://www.mpa-garching.mpg.de/integral/survey/catalog.php} and \cite{bird10}. An up to date collection of all the sources discovered by \textit{INTEGRAL} can be found at the \textit{INTEGRAL} IGR Source page\footnote{http://irfu.cea.fr/Sap/IGR-Sources/}. Last but not least, the ISDC has developed and is maintaining an all \integral~instrument archive, known as HEAVENS \footnote{http://www.isdc.unige.ch/heavens/} that collects light-curves, images and spectra for a given source or position in the sky, from the public data.

Notwithstanding all the above, we believe that having a complementary local archive of \integral~data and  products  that is easy to access and to handle (be it for only one instrument) is an important way to boost scientific knowledge. It is open to suggestions and ideas of the local users in a close and frequent feedback process that maximizes the 
scientific outcome. 
Developing and maintaining a  database enables complete control over what is being done and how; raw data and products are accessible at all levels (i.e. analysis steps), entire sky maps are available (not just a catalogue or a zoom on the target, but complete \textit{workable} FITS files) and the results of \textit{all} the detected sources can be retrieved very quickly, regardless the time interval: e.g. long term light-curves can be easily extracted also by non-\integral~experts (a matter of a few seconds for the 2\,ks binning).
A systematic overview of different classes of sources can be performed, stimulating  new investigations and collaborations, as shown in Section~\ref{sec:results}, 
and enabling interesting discoveries,  determination of trends, as well as providing  a complementary picture to other on-going missions.

\section{Acknowledgements}
Based on observations with \textit{INTEGRAL}, an ESA project
with instruments and science data centre funded by ESA member states
(especially the PI countries: Denmark, France, Germany, Italy,
Spain, and Switzerland), Czech Republic and Poland, and with the
participation of Russia and the USA. AP thanks S.E.~Shaw, B.E.~O'Neel and the ISDC staff since most of the know-how needed to 
set-up such an archive was gained during her years at the ISDC (2000-2005).
 We acknowledge the Italian Space Agency financial  support
via contract {\it INTEGRAL}  ASI-INAF I/033/10/0. We thank the anonymous Referee for his/her comments and suggestions, which improved the quality of the paper.

\bibliographystyle{model2-names}
\bibliography{biblio}







\end{document}